\documentclass[12pt,preprint]{aastex}
\usepackage{emulateapj5}
\usepackage{graphicx}

\def\xte{{\it RXTE}}

\def\asca{{\it ASCA}}
\def\chandra{{\it Chandra}}

\def\xray{\hbox{X-ray}}
\def\nh{\hbox{$N_{\rm H}$}}

\def\spn{\rm{PWN}}
\def\pwn{\rm{PWN}}
\def\psr{\rm{PSR}}

\slugcomment{To appear in the June 20, 2003 issue of the Astrophysical Journal}

\shortauthors{Gotthelf}
\shorttitle{X-ray Spectra of Young Pulsars and Their Nebulae: Spin-down Energy}

\begin{document}

\title{X-ray Spectra of Young Pulsars and Their Wind Nebulae:\\ Dependence on Spin-down Energy}

\author{E. V. Gotthelf}
\affil{Columbia Astrophysics Laboratory, Columbia University, 550 West 120$^{th}$ Street, New York, NY 10027, USA; evg@astro.columbia.edu}

\begin{abstract}

An observational model is presented for the spectra of young
rotation-powered pulsars and their nebulae based on a study of nine
bright Crab-like pulsar systems observed with the {\it Chandra} \xray\
observatory.  A significant correlation is discovered between the
\xray\ spectra of these pulsars (\psr s) and that of their associated
pulsar wind nebulae (\spn e), both of which are observed to be a
function of the spin down energy, $\dot E$. The $2-10$ keV spectra of
these objects are well characterized by an absorbed power-law model
with photon indices, $\Gamma$, in the range of $0.6 < \Gamma_{\psr} <
2.1$ and $1.3 < \Gamma_{\spn} < 2.3$, for the pulsars and their
nebulae, respectively. A linear regression fit relating these two sets
of indexes yields, $\Gamma_{\spn} = 0.86\pm 0.20 + (0.72\pm 0.13)
\times \Gamma_{\psr}$, with a correlation coefficient of $r=0.96$.
The spectra of these pulsars are found to steepen as $\Gamma =
\Gamma_{max} + \alpha \dot E^{-1/2}$, with $\Gamma_{max}$ providing an
observational limit on the spectral slopes of young rotation-powered
pulsars.  These results reveal basic properties of young pulsar
systems, allow new observational constraints on models of pulsar wind
emission, and provide a means of predicting the energetics of such
systems when lacking detected pulsations.

\end{abstract}

\section{Introduction }

The classical picture of a pulsar wind nebula (\pwn) is that of a
bright, centrally condensed, diffuse nebula whose broad-band
non-thermal continuum spectrum is characterized by a power-law with
one or more breaks (see Chevalier 1998 and Arons 2002 for reviews of
\pwn\ models). Recent observations of these objects with the \chandra\
\xray\ observatory has resolved out complex structure involving
co-aligned toroidal arcs, axial jets, and wisps on arc-second scales
(e.g., Crab Nebula, Weisskopf et al. 2000; Vela Nebula, Helfand et
al. 2000; 2001). This basic morphology appears to be common among
young, energetic pulsars associated with supernova remnants (Gotthelf
2001). Furthermore, \chandra\ imaging-spectroscopy in the $2-10$ keV
band reveals a distinct spectral signature between these structures and
that of the central pulsar, blended in earlier studies (Gotthelf \& Olbert
2002). It is now clear that current models for the classic \pwn\ are
inadequate in describing their observed structure or spectrum.

This Letter presents a spectral analysis of nine pulsars observed by
\chandra\ which show bright, resolved \pwn e. The high resolution
\chandra\ data allows for the first time a systematic \xray\ spectral
study of pulsars and their nebulae independently.  For the $2-10$ keV
energy band, the spectral slopes of these pulsars and their associated
nebulae are found to be tightly correlated over the full range of
measured values implying a fundamental spectral relationship not
easily explained in the framework of current models of pulsar
emission. The spectral slopes are also shown to follow an inverse
square-root relation with respect to the spin-down energy - the spectral
slope of younger, more energetic pulsars tend to be steeper,
suggesting that the physics of a pulsar's particle acceleration is
governed primarily by its spin-down energy.

Collectively, these results provide new observational insight into the
emission mechanisms of young rotation-powered neutron stars,
characterize a pulsar's spin-down evolution, and strongly constrains
pulsar emission models. 

\section{NEW CHANDRA OBSERVATIONS}

The \chandra\ \xray\ Observatory (Weisskopf, O'Dell, \&\ van
Speybroeck 1996) has targeted most pulsars with known \xray\ bright
wind nebulae. Table~1 presents the sample used in this study. Each
pulsar was imaged by the Advanced CCD Imaging Spectrometer (ACIS)
whose arc-second resolution allows us to isolate the pulsar emission
from that of the nebula. ACIS is sensitive to \xray s in the
0.2--10~keV spectral band with an energy resolution of $\Delta E / E
\sim 0.1$ at $1$ keV. All data were corrected for CTI effects by
reprocessing the Level 1 data and applying the standard Level 2
filtering criteria following the Townsley method (Townsley et
al. 2000). Time intervals of anomalous background rates associated
with particle flare events were further rejected.  Data reduction and
analysis were accomplished using the CIAO software package (CIAO
2.2/CALDB v.2.9).

For each object, we extracted spectra from the pulsar, nebula, and their
local backgrounds, when available, or obtained spectral parameters
from the literature as noted in Table~1. Pulsar extraction regions are based
on the image above 4.0\,keV, where the count rate per pixel from the
pulsar dominates over the surrounding diffuse nebula emission.  Large
variations in nebula size and background rates for the sample
precluded the use of a standard extraction aperture. Instead, the
nebula region was defined by the $3\sigma$ contour above the
background level with the pulsar region excluded. All spectra were
grouped to a minimum of 50 counts per spectral bin and model spectra
fitted using the {\tt XSPEC} spectral fitting package (version v11.1).
The ACIS instrument response matrix provided with the Townsley CTI correction
software were used in these fits, along with an
ancillary telescope response function created according to standard
CIAO 2.2 procedures, using the Townsley QEU calibration file. 

An important consideration in our analysis is spectral distortion due
to CCD photon pile-up. Pile-up occurs when more than one photon is
recorded as a single photon event in a CCD read-out frame. This can
severely skew the spectrum by registering the summed energy of two
photons as one; it also causes the loss of soft photons when the
recorded energy of two or more photons exceed the valid range of the
instrument. For our measured pulsars, the expected magnitude of
pile-up spans the range of $3-8\%$. Pile-up can result in the spectral
flattening of the pulsar spectra relative to that of the nebulae and
is accounted for using the pile-up model available in {\tt XSPEC} on
all pulsar spectral fits.

The extracted spectra were fitted with an absorbed power-law model
whose flux is given by $F(E) \propto e^{-\sigma(E)\nh} E^{-{\Gamma}}$,
were $\Gamma$ is the photon index, \nh\ the interstellar hydrogen
column density to the source, and $\sigma(E)$ the Wisconsin
interstellar absorption cross-section (Morrison \& McCammon 1983). In
order to avoid potential soft thermal emission contamination, the
spectral fits for both the pulsars and nebulae were restricted to
energies above $2$~keV, where the fits are largely insensitive to \nh.
In these fits, the \nh\ was held fixed to the value determined from
fits to the nebula spectra over the full ACIS band. The spectral fits
to the pulsar emission included the addition of a CCD pile-up model.

Table~1 lists the resultant spectral slopes for each object, including
the spectra slope of the {\it pulsed} emission only, taken from
published values, when available, or derived by phase-resolved
spectroscopy using \asca\ or \xte\ data.  A comparison between pulsed
and unpulsed emission measurements is a good test for systematic
spectral distortion, as the former measurements, based on
phase-resolved spectroscopy using non-CCD detectors, are immune to
pile-up. Although the spectra of the pulsed and total emission need
not be the same (e.g., Crab; see Pravdo et al. 1997), it is reassuring
that most agree to within measurement errors (see Figure 1, Gotthelf
2003). Interesting new exceptions are the N157B and 3C58 pulsars which
predict a phase dependence in their spectral slopes.

\section{RESULTS}

A comparison of the spectral indices and three interesting
parameters among the objects shown in Table~1 reveals some remarkable
trends. A plot of the index for each pulsar versus that of its
structured nebula is shown in Figure 1.  A linear regression fit to
this data, taking into account the uncertainties in both coordinates,
yields,

\begin{equation} 
 \Gamma_{\spn} =  0.86 (0.20) + 0.72 (0.13) \times \Gamma_{\psr}
\end{equation} 

\noindent where the one-sigma errors are given in parentheses. This
relationship is highly significant, with a linear correlation
coefficient of $r = 0.96$, and covers the range of known pulsar
spectral power-law indexes in the $2-10$ keV \xray\ band. This result
is rather surprising, since no such correlation was previously known
or predicted.

Another unexpected result is found by comparing the rank-ordered
pulsar indices with the spin-down energies of the associated object
(see Table 1). Allowing for the uncertainties in $\Gamma_{\psr}$, an
unambiguous trend is evident, with a clear spectral steepening with
spin-down energy, $\dot E$. Since the spin-down energy is expected to
play an important role in the evolution of the pulsar and its nebula,
various functional forms involving the spin-down energy were examined
in order to model the spectral slope. The best fit is obtained with an
inverse square root model, $\Gamma = \Gamma_{max} + \alpha \dot
E^{-1/2}$, with the following parameters (see Figure 2):

\begin{equation} 
\Gamma_{\psr} = 2.08 (0.07) - 0.029 (0.003) \ \dot E_{40}^{-1/2}
\end{equation} 

\noindent where $\dot E_{40}$ is the spin-down energy in units of $
10^{40}$ erg s$^{-1}$ and the one-sigma errors are given in
parentheses. The fit is rather poor with $\chi^2 = 17$ for $7$ DoF,
but nearly all the excess contribution to $\chi^2$ is from the Kes 75
data point, whose $\dot E$ has been previously noted to be at odds
with the properties of the SNR (Helfand, Collins, Gotthelf
2002). Without this data point, the fit is excellent ($\chi^2 = 5$ for
6 DoF), however the model parameters are not significantly changed and
we quote the fit using all nine data points.

Given Equation 1, we expect $\Gamma_{\spn}$ to follow a similar
functional form. This is found to be the case but with a lesser
significance of $r = 0.90$ ($\chi^2 = 54$ for $7$ DoF), again with the
Kes 75 point providing the greatest contribution to $\chi^2$. We thus
use Equations 1 and 2 to derive the complementary equation for the
spectral slopes of the structured nebula:

\begin{equation} 
\Gamma_{\spn} = 2.36 (0.33) - 0.021 (0.005) \ \dot E_{40}^{-1/2}
\end{equation} 

The above two equations, along with the updated Seward \& Wang (1988)
$L_x / \dot E$ relationship (Possenti et al. 2002), relate the timing
properties of a rotation-powered pulsar to its X-ray spectra in the
$2-10$ keV energy band.  These equations predict a maximum value for
the spectral slopes, $\Gamma_{max}$, for pulsars obeying these
equation. Furthermore, in lieu of any high-energy spectral cut-off,
energetic considerations restrict the minimum spectral slopes, and
therefore $\dot E$, to $0 < \Gamma_{\psr} < 2.2$ and $ 0 <
\Gamma_{\spn} < 2.7$. Additional emission-model dependent restrictions
(e.g., putative particle distribution spectral slope; see \S 4) may
further constrain the lower limit on $\Gamma$.

Figure 3 shows the permitted range of $\Gamma_{\psr}$ in the $P-\dot
P$ diagram for a young rotation-powered pulsar. Radio pulsars which
lie in this region, but not already in our sample, likely contain
undetected nebulae whose spectral slopes can be predicted. This
population may also offers a mean to determine the $\Gamma_{\psr}$
cut-off for Crab-like pulsars. The former is a test of the hypothesis
that observable \spn\ are restricted within the $P-\dot P$ plane and
the latter will provide an important constraint on pulsar emission
models.

A trend is also noticed for the spectral hardness in the $2-10$ keV
band which decreases with characteristic age $\tau$ (see Table 1), but
unlike for the spin-down energy, the correspondence is not
consistently monotonic. Considering the ordering by spin-down energy,
large deviations by an individual pulsar from the trend may be a good
indicator that $\tau$ is in fact a poor measure of that pulsar's true
age. And finally, the magnetic field is also expected to play an
important role in the evolution of a \spn, particular in determining
the synchrotron lifetime of accelerated particles (see \S 4).
However, no simple trend is found for the derived magnetic field with
respect to the measured spectral slopes or to the other calculated
parameters.

\section{DISCUSSION}

The broadband spectra of the Crab pulsar and nebula have been known
for some time (e.g. see Fig. 4-2 of Manchester \& Taylor 1977) and
agrees well with the spectral relationship presented herein.
But until now, no systematic correlation linking a pulsar's radiant
spectrum with that of its nebula or to its spin-down energy was either
noted or predicted.  A key question is where does the pulsar emission
occur and how is it related to the observed structured nebula; this is
consider below in the context of current theory.

Both competing models for generating high-energy emission from a
pulsar, the polar-cap (e.g., Harding et al. 1978) and outer-gap (e.g.,
Cheng, Ho, \& Ruderman 1986) models, provide a natural particle
accelerator, radiating synchrotron emission close to the star (within
the light cylinder) in a strong magnetic field.  The standard theory
of synchrotron acceleration of particles with large pitch angles
($\Phi_{pitch} >> 1/\gamma$; $\gamma$ is the Lorentz factor) predicts
a radiation spectrum of $I_{\nu} \propto\ \nu^{-(p-1)/2}$ for a
particle distribution $N(\gamma) = N_o\gamma^{-p}$, radiating away
their energy on the order of seconds. Most models estimate and/or
simulation $p \simeq 2$ (e.g., Crusiu-W\"atzel, Kunzl, \& Lesch
2001). Ultimately, in a pure synchrotron model, energy constraints
require a particle index $p$ ($= 2\Gamma - 1$) greater than 2, forcing
a lower bound on $\Gamma >2.0$, above the synchrotron cooling
frequency, or $\Gamma > 1.5$, below, generally in the X-ray regime.
This is problematic for these models given the range of
$\Gamma_{\psr}$ presented in Table~1.  The addition of a curvature
radiation component and the location of the cyclotron turnover may
allow for spectral slope variations in the $2-10$ keV band (e.g.,
Rudak \& Dyks 1999). But how this might be connected to a pulsar's
spin-down energy is unclear.

In contrast to the highly localized accelerator in the above pulsar
emission models, the standard Kennel \& Coroniti (1984a,b) model for
the Crab nebula is based on a symmetric, isotropic wind.  In this
model, a young pulsar loses its rotational energy predominantly in the
form of a highly relativistic ($\gamma \sim 10^6$), isotropic particle
wind.  The freely expanding wind is initially invisible as it travels
though the surrounding self-evacuated region, but eventually
encounters the ambient medium ($r_s \sim 0.1$ pc) where it is
reverse-shocked, resulting in the redistribution of particle energies
with that of the magnetic field. The visible nebula is manifest as the
shocked particle emitting synchrotron radiation. The above symmetric
model is, however, notably inadequate in explaining the observed tori
and jet-like structure.

In the context of this work, these models do not provide a physical
mechanism to link the pulsar and nebula spectrum. Our results suggest
that the structured nebula is either powered directly by the pulsar or
both are responding to a common mediating process. Although the
particles associated with the pulsed emission are available for the
shocked wind, they are not expected to retain their original
distribution and track the pulsar's spectrum. In fact, the Kennel \&
Coroniti (1984a,b) model for the Crab is inconsistent with a particle
wind originating from the pulsar based on energetic grounds. A wind
with a time-average injection rate of $\dot N_{\pm} \sim 3 \times
10^{40}$ for $e^{\pm}$ pairs with $\gamma \sim 10^6$ would exceed the
spin-down power of the Crab by several orders of magnitude (Gallant et
al. 2001). Furthermore, in the $2-10$ keV band, Fermi acceleration of
highly relativistic particles yield injected spectrum with fixed $p
\sim 2.2-2.3$ (Achterbers et al. 2001). Both the range of observed
\spn\ slopes and their correlation with the \psr\ emission strongly
constrain all models involving shock acceleration as the origin of the
particle distribution. Investigations of collisionless models may be
fruitful (e.g. Blanford 2003).

The apparent steepening of the spectral slope with spin-down energy
suggests that the spectral evolution is best parameterized by $\dot E$
instead of the characteristic age $\tau$. Of all the derived
quantities, $\dot E$ is the least model dependent, likely to be an
accurate measure of the current energy loss rate. $\tau$ is
evidentially an uncertain estimate of a pulsar's true age and it is
less well correlated with spectral slope. Young pulsars are known to
undergo episodes of glitches, and both internal and external torques
can adjust their period and period derivatives.
This allows for the possibility of distinct episodes of increased
efficiency in the $\dot E$ energy loss during a pulsars' early
evolution.  This is might be consistent with the relatively large
characteristic age for the bright, diffuse nebulae of N157b and 3C58,
perhaps currently in transition.

The \spn\ of some pulsars show a gradual spectral steepening away from
the center of the nebulae, most notably for the Crab and 3C58
nebula. This is generally attributed to synchrotron losses from a
centrally injected power-law distribution particle wind.  If the
synchrotron loss time is less than the pulsar's age, a steepening of
$\Delta\Gamma = 0.5$ in the spectrum is expected across the
nebula. Observationally, Equation 1 rules out a simple synchrotron
cooling mechanism (i.e., $\Delta\Gamma = \Gamma_{\spn} - \Gamma_{\psr}
= -0.028 \ \Gamma_{\spn} + 0.86$). However, because of the relatively
large uncertainties on the spectral indices, a $\Delta\Gamma = 0.5$
solution is not formally excluded, as the fit for the latter model is
only marginally worse then for the former ($\chi^2 = 8$ vs. $5$, for 7
DoF, respectively). Deeper \chandra\ observations, with reduced
measurement errors, are needed to distinguish fits.


The well defined spectral relationship for our pulsar sample suggests
that the observational properties of Vela as compared to the Crab are
simply the consequence of an older, less energetic system. It is also
apparent that the manifestation of a shell remnant associated with a
\psr/\spn\ system, the main observational difference between Vela and
the Crab, must be to some degree autonomous of either pulsar or SNR
evolution.  This raises questions for the importance of a pressure
confined wind, inherent in \spn\ evolution models, to account for the
observed morphology of the structured nebula.

Considering that \chandra\ is the first telescope to spatially resolve
neutron stars from their structured nebulae, previous spectral
observations of young pulsars are likely to be contaminated by nebula
emission.  For example, we see no clear relationship between spectral
indices of pulsars obtained by {\it ASCA} with those measured by
{\it Chandra}.  Fits to these curved composite spectra using single
absorbed power-law models, typically assumes, can result in
exaggerated absorption column measurements.

For pulsars lacking measured pulsations, the relationships presented
herein provide a powerful tool for predicting the spin-down energy of
pulsars directly from their spectra. For an unresolved pulsar (and
nebula) the presence of a nebula can also be predicted, based on the
measured spectrum. As more young rotation-powered pulsars are observed
with \chandra, a more complete picture will emerge of the
morphological and spectral characteristics of young pulsar evolution,
paving the way for future, perhaps unified, models of pulsars and
their structured nebulae.

\section{ACKNOWLEDGMENTS}

This work is made possible by NASA LTSA grant NAG~5-7935. I thank
Charles M. Olbert and Benjamin F. Collins for assistance with the
presented spectral analysis. I thank Jules P. Halpern for invaluable
discussions.


\begin{deluxetable}{llcccccc}
\tabletypesize{\footnotesize}
\tablewidth{0pt}
\tablecolumns{7}
\tablecaption{X-ray Spectra of Young Pulsars with Bright Central Nebulae Observed with {\it Chandra}$^a$}
\tablehead{
\colhead{Remnant} & \colhead{Pulsar} & \colhead{$\Gamma_{\spn}^{b}$} & 
\colhead{ $\Gamma_{\psr}^{c}$}    & \colhead{$\Gamma_{\rm Pulsed}^{d}$} & 
\colhead{ {$\tau^{e}$} } & \colhead{ {$\dot {\rm E}^{e}$} }  & \colhead{ {B$_p$/B$_{\rm{QED}}^e$} } \\
 \colhead{} &  \colhead{} &  \colhead{} &  \colhead{} &  \colhead{}  &  
\colhead{(kyr)} &  \colhead{ ($\times 10^{35}$ ergs/s) } & \colhead{}
}
\startdata
G11.2$-$0.3 & PSR~J1811$-$1926 & 1.28$\pm$0.15 & 0.63$\pm$0.12 & 0.60$\pm$0.60  &23.0  &   53  &0.04\\
Vela~XYZ    & PSR~J0835$-$4510 & 1.50$\pm$0.04 & 0.95$\pm$0.24 & 0.93$\pm$0.26  &12.0  &   67  &0.08\\
G54.1$+$0.3 & PSR~J1930$+$1852 & 1.64$\pm$0.18 & 1.09$\pm$0.09 & 1.06$\pm$0.86  &2.9   &  118  &0.23\\
Kes~75      & PSR~J1846$-$0258 & 1.92$\pm$0.04 & 1.39$\pm$0.11 & 1.10$\pm$0.30  &0.7   &   82  &1.10\\
MSH~15$-$52 & PSR~J1513$-$5908 & 1.93$\pm$0.03 & 1.40$\pm$0.50 & 1.26$\pm$0.08  &1.6   &  140  &0.34\\
3C\,58      & PSR~J0205$+$6449 & 1.92$\pm$0.11 & 1.73$\pm$0.15 & 1.11$\pm$0.34  &5.0   &  263  &0.08\\
SNR~0540-69 & PSR~J0540$-$6919 & 2.09$\pm$0.11 & 1.88$\pm$0.11 & 1.83$\pm$0.13  &1.7   & 1481  &0.11\\
Crab~Nebula & PSR~J0534$+$2200 & 2.14$\pm$0.01 & 1.85$\pm$0.09 & 1.87$\pm$0.05  &1.3   & 4394  &0.08\\
N157B~Neb.  & PSR~J0537$-$6910 & 2.28$\pm$0.12 & 2.07$\pm$0.21 & 1.60$\pm$0.35  &5.0   & 4916  &0.02\\
\enddata
\tablenotetext{a}{Ranked by increasing pulsar photon index, $\Gamma_{\spn}$ (then $\Gamma_{\psr}$, were ambiguous). Data for the Crab pulsar and nebula were obtained from Pravdo, Angelini \& Harding (1997). }
\tablenotetext{b}{Values for the following objects were taken from the literature: G54.1$+$0.3: Lu et al. 2002; SNR~0540-69: Kaaret et al.~(2001).}
\tablenotetext{c}{Includes both pulsed and un-pulsed emission, corrected for pile-up.}
\tablenotetext{d}{Pulsed spectrum references: Vela: Strickman, Harding \& Jager (1999); G11.2-0.3: Torii et al.~(1997); Kes~75: Gotthelf et al.~(2000); N157B: Marshall et al.~(1998); SNR~0540-69: Kaaret et al.~(2001).}
\tablenotetext{e}{The characteristic pulsar spin-down age is defined as  $\tau \equiv\ P/2\dot P$, the spin-down energy as, $\dot E = 4\pi^2 I \dot P/P^3$, were $I \equiv 10^{45}$ gm cm$^{-2}$, and the  inferred magnetic field B$_{p} = -6c^3{\dot E} /R^6\Omega^4$ G is normalized to the quantum critical field, B$_{QED} = m^2_e c^3 / e \hbar = 4.4 \times 10^{13}\ {\rm G}$.}
\end{deluxetable}




\begin{figure}
\begin{center}
\includegraphics[width=75mm]{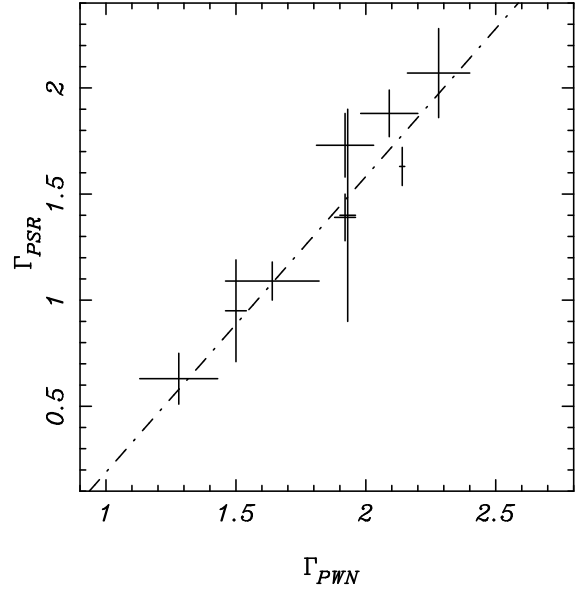}
\caption{Relationship between a pulsar's spectral slope
($\Gamma_{\psr}$) and that of its structured nebulae ($\Gamma_{\spn}$)
in the $2-10$ keV energy range, assuming a simple power-law model for
the objects presented in Table~1. The dashed-line indicates the
best-fit linear model. The physical origin of this relationship has
yet to be determined}
\end{center}
\end{figure}


\begin{figure}
\begin{center}
\includegraphics[width=150mm]{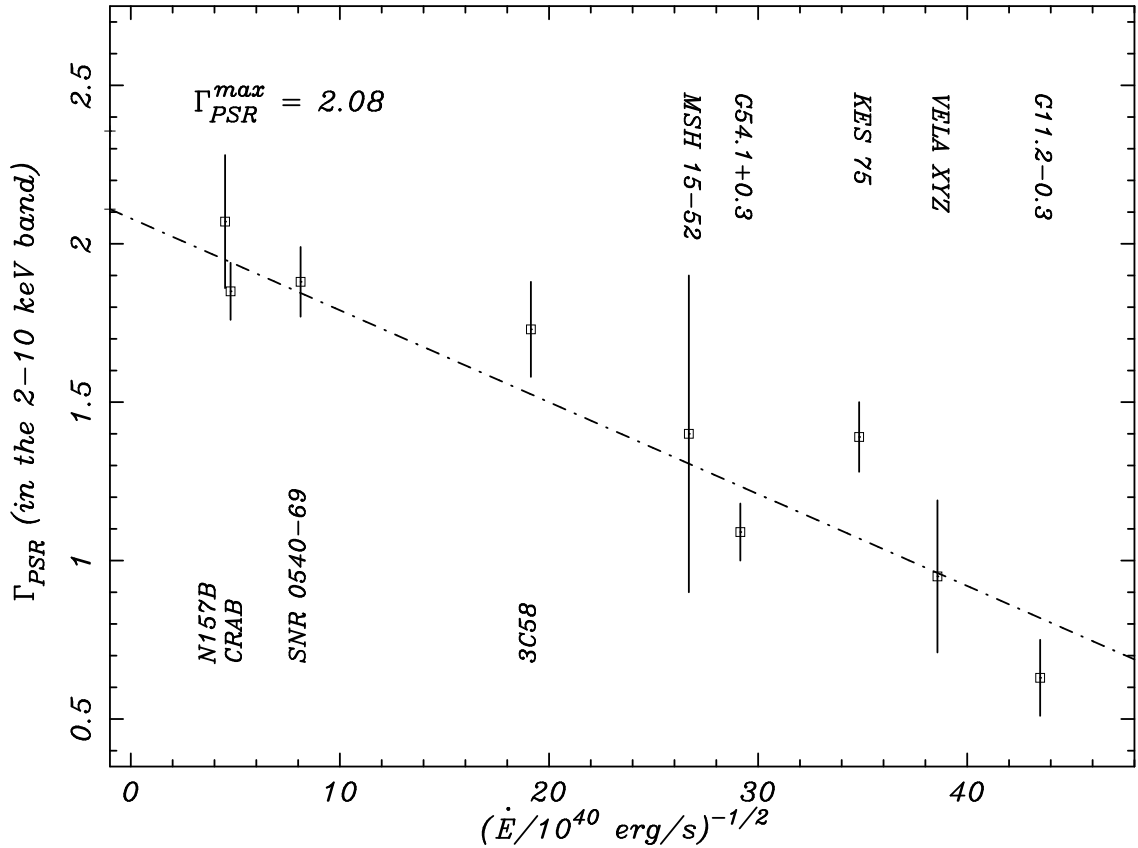}
\caption{A comparison between the spectral slope of the $2-10$ keV
pulsar emission ($\Gamma_{\psr}$) and the square root of the spin-down
energy, $\dot E_{40}^{-1/2}$, in units of $10^{40}$ erg s$^{-1}$, for
the each object presented in Table~1. The dashed-line indicates the
best-fit model.}
\end{center}
\end{figure}

\begin{figure}
\begin{center}
\includegraphics[width=150mm]{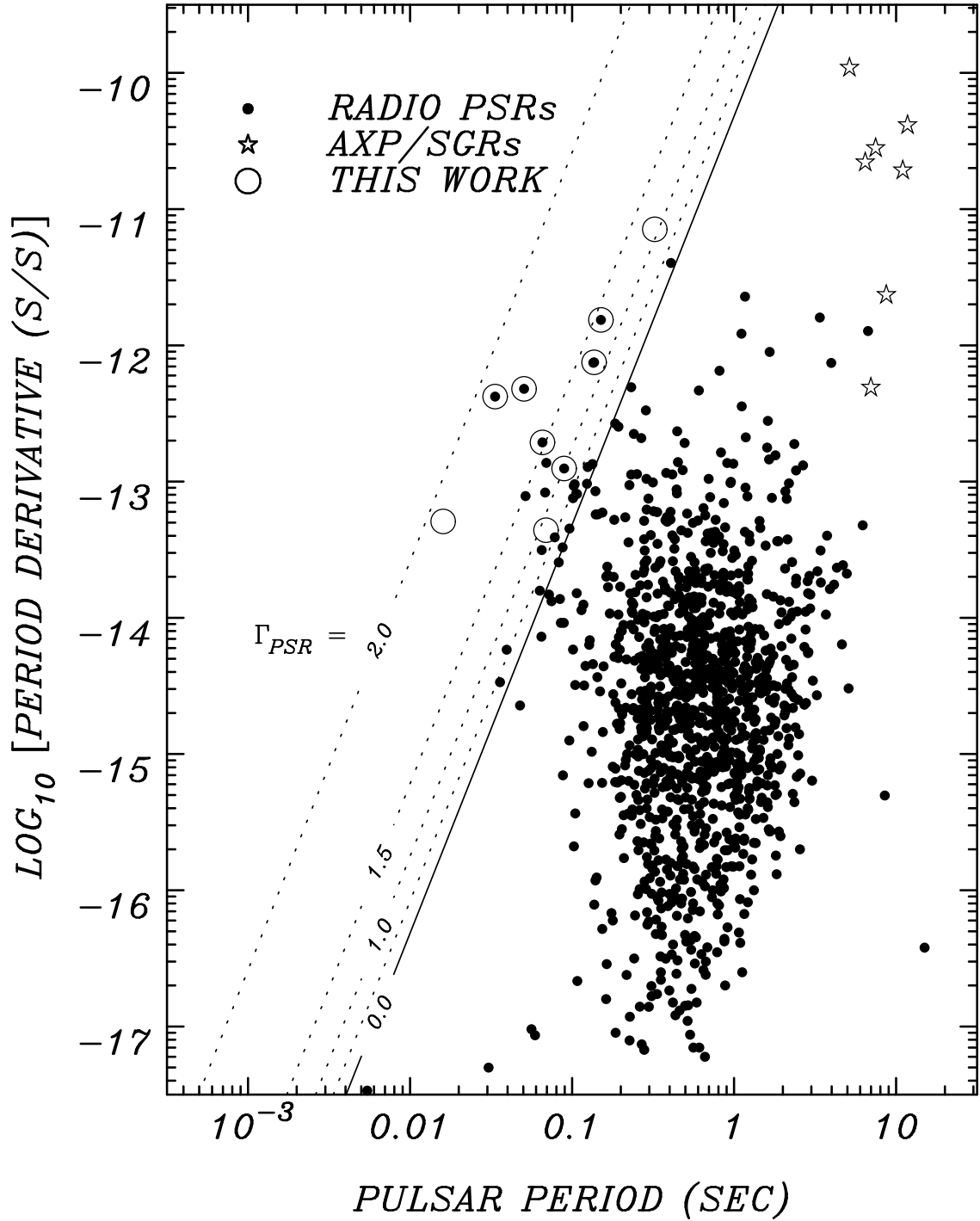}
\caption{The period-period derivative diagram showing the sample of
X-ray objects presented in Table~1 (open circles), a complete sample
of known radio pulsars from Manchester (2003) (filled circles), and
the known AXPs/SGRs (stars). Lines of constant spectral slope for a
young, rotation-powered pulsar (in the $2-10$ keV energy band; see
text) is given by the diagonal lines. The solid line corresponds to
$\Gamma_{\psr} = 0$. }
\end{center}
\end{figure}


\begin{references}

\reference{} Achterberg, A., Gallant, Y.~A., Kirk, J.~G., Guthmann, A.~W. 2001, MNRAS, 328, 393
\reference{} Arons, 2002 in ``Neutron Stars in Supernova Remnants'', ASP Conf. Ser., 271, 71
\reference{} Blandford, R., D. 2003, ``High energy Processses and Phenomina in Astrophysics'', IAU Symp No. 214, in press
\reference{} Cheng, K.~S., Ho, C., \& Ruderman, M. 1986, ApJ, 300, 500
\reference{} Chevalier, R. A. 1998, MmSAI, 69, 977
\reference{} Crusiu-W\"atzel, Kunzl, \& Lesch 2001, ApJ, 546, 401
\reference{} Kennel, D.~F. \& Coroniti, F. V. 1984a, 283, 694
\reference{} Kennel, D.~F. \& Coroniti, F. V. 1984b, 283, 710
\reference{} Harding, A.~K., Tademaru, E., \& Esposito, L. S. 1978, ApJ, 225, 226
\reference{} Helfand, D.~J., Gotthelf, E. V., \& Halpern, J. P. 2001, ApJ, 556, 380
\reference{} Helfand, D.~J., 2000, HEAD, 32, 3603
\reference{} Hoshino, M.,  Arons, J,, Gallant, Y. A., Langdon, A. B. 1992, ApJ, 390, 454
\reference{} Gotthelf, E.~V., et. al. 2000, \apjl, 542, L37
\reference{} Gotthelf, E.~V. 2001, AIP Conf. Proc., Vol. 586, p. 513
\reference{} Gotthelf, E.~V. \& Olbert, C.~M. 2002, in ``Neutron Stars in Supernova Remnants'', ASP Conf. Ser. 271, 171.
\reference{} Gotthelf, E.~V. 2003, in ``High Energy Processes and Phenomena in Astrophycs'', IAU Symposium No. 214, in press
\reference{} Kaaret, P., et. al 2001, ApJ, 546, 1159
\reference{} Kaspi, V.~M., et al. 2001, ApJ, 560, 371
\reference{} Lu, F.~J., et al. 2002 ApJ, 568, 49
\reference{} Manchester, R. N. \& Taylor, J. H. 1977, ``Pulsars'', W. H. Freeman, San Fran. QB843.P8 M3
\reference{} Marshall, F.~E., et al. 1998, \apjl, 499, L179
\reference{} Morrison, R. \& McCammon, D. 1983, ApJ, 270, 119
\reference{} Mereghetti, S., Bandiera, R., Bocchino, F., Israel, G. L. 2002, ApJ, 574, 873
\reference{} Pravdo,  S.~H., Angelini, L., \& Harding, A.~K.\ 1997, \apj, 491, 808 
\reference{} Rudak, B. \& Dyks, J. 1999, MNRAS, 303, 477
\reference{} Strickman, M.~S., Harding, A.~K., \& de Jager, O.~C.\ 1999, \apj, 524, 373
\reference{} Townsley, L.~K., Broos, P.~S, Garmire, G.~P., Nousek, J.~A., 2000, \apjl, 534, L139
\reference{} Torii, K., Tsunemi, H., Dotani, T.,\&  Mitsuda, K. 1997 ApJ, 489, L145
\reference{} Torii, K., Tsunemi, H.,  Dotani, T, Mitsuda, K., Kawai, N., Kinugasa, K., Saito, Y., Shibata, S. 1999, ApJ, 523, 69
\reference{} Weisskopf, M.~C., et al. 1996, SPIE 2805, III, 2
\reference{} Weisskopf, M.~C., et al. 2000, ApJ, 536, L81

\end{references}
\end{document}